\documentclass[journal=nalefd, manuscript=letter, layout=twocolumn]{achemso}
\usepackage[version=3]{mhchem} % Formula subscripts using \ce{}
\usepackage{color}
\usepackage{braket}
\usepackage{subcaption}
\usepackage{siunitx}
\usepackage{xcolor}
\usepackage{graphicx}
\usepackage{graphbox}

\usepackage[colorlinks=true,
                linkcolor=blue,
	        citecolor=blue,
	        urlcolor=blue]{hyperref}

\newcommand{\Ex}{E_{\rm{ext}}}
\newcommand{\Ei}{E_{\rm{int}}}

\newcommand{\theos}{Theory and Simulation of Materials (THEOS), and National Centre for Computational Design and Discovery of Novel Materials (MARVEL), \'Ecole Polytechnique F\'ed\'erale de Lausanne, CH-1015 Lausanne, Switzerland}
\newcommand{\trieste}{Dipartimento di Fisica, Universit\`a di Trieste, Strada Costiera 11, 34151 Trieste, Italy}
\newcommand{\liege}{nanomat/QMAT/CESAM and European Theoretical Spectroscopy Facility
, University of Liège, Allée du 6 Août 19 (B5a), 4000 Liège, Belgium}

\newcommand\equally{These authors contributed equally to this work.}

\title{Gate control of spin-layer-locking FETs and application to monolayer LuIO}

\author{Rong Zhang}
\altaffiliation{\equally}
\affiliation{\theos}
\author{Antimo Marrazzo}
\altaffiliation{\equally}
\affiliation{\theos}
\alsoaffiliation{\trieste}
\author{Matthieu Verstraete}
\affiliation{\liege}
\author{Nicola Marzari}
\affiliation{\theos}
\author{Thibault Sohier}
\email{thibault.sohier@uliege.be}
\affiliation{\liege}
\alsoaffiliation{\theos}
\begin{document}

\begin{abstract}
A recent 2D spinFET concept proposes to switch electrostatically between two separate sublayers with strong and opposite intrinsic Rashba effects. This concept exploits the spin-layer locking mechanism present in centrosymmetric materials with local dipole fields, where a weak electric field can easily manipulate just one of the spin channels.
Here, we propose a novel monolayer material within this family, lutetium oxide iodide (LuIO). It displays one of the largest Rashba effects among 2D materials (up to $k_R = 0.08 \si{\angstrom}^{-1}$), leading to a $\pi/2$ rotation of the spins over just 1 nm.
The monolayer had been predicted to be exfoliable from its experimentally-known 3D bulk counterpart, with a binding energy even lower than graphene.
We characterize and model with first-principles simulations the interplay of the two gate-controlled parameters for such devices: doping and spin channel selection. We show that the ability to split the spin channels in energy diminishes with doping, leading to specific gate-operation guidelines that can apply to all devices based on spin-layer locking.
\end{abstract}

%\section{Introduction}
\textbf{Introduction.} Spintronics aims at improving the efficiency of electronic devices and to enrich them with new functionalities, ultimately  delivering multi-functional, high-speed, low-energy electronic technologies \cite{hu_recent_2020}.
In spintronic devices the information is encoded into the electronic spin state, which can be manipulated and transported similarly to the electronic charge in conventional electronics.
In 1990 \cite{datta_electronic_1990}, Datta and Das proposed a field-effect transistor based on the electron spin (spinFETs) that could potentially operate at low power and provide high computing speed.
The Datta-Das spinFET consists of a two-dimensional electron gas (2DEG) with Rashba spin-orbit coupling (SOC), as can be realized in narrow-gap semiconductors, like InGaAs/InAlAs heterostructures \cite{datta_electronic_1990} or two-dimensional (2D) materials \cite{hu_recent_2020}, that are placed between ferromagnetic contacts and under an electrical gate.
The input and output contacts have orthogonal directions of magnetization, such that only electrons with precessing spin can be collected by the drain.
The Rashba SOC is responsible for a spin precession such that a $\pi/2$ phase rotation can in principle be achieved across distances that are shorter than the mean free paths of high-mobility semiconductors at low temperatures \cite{datta_electronic_1990}.
In this field-effect setup, the magnitude of the Rashba SOC---and so the spin precession---is modulated through an electrical gate.

A large Rashba SOC is needed to achieve short precession lengths, which can be found in compounds that contain heavy chemical elements and exhibit a non-centrosymmetric crystal structure (hereafter called R-1 materials \cite{zhang_hidden_2014}), where crystal inversion symmetry is globally broken.
However, in these materials the gate voltage usually induces relatively weak external electric fields compared to the internal field due to the broken symmetry, resulting in a weak modulation of the Rashba SOC.
Hence, the traditional Datta-Das model faces substantial challenges in the manipulation and reversal of the electron spin by a gate voltage.
A possible remedy to this drawback has been proposed in Ref. \citenum{LaOBiS2NanoLetter},  based on using centrosymmetric 2D materials that break inversion symmetry only locally and not globally.
Refs. \citenum{zhang_hidden_2014, riley_direct_2014} showed how Rashba and Dresselhaus SOC can emerge also in centrosymmetric materials, provided that the inversion symmetry is locally broken on atomic sites, owing to the local nature of the SOC effect.
Following the convention of Ref. \citenum{zhang_hidden_2014} these centrosymmetric materials with local dipole fields and Rashba SOC are called R-2 materials.
\begin{figure*}
  \centering
  \includegraphics[align=c,width=0.49\textwidth]{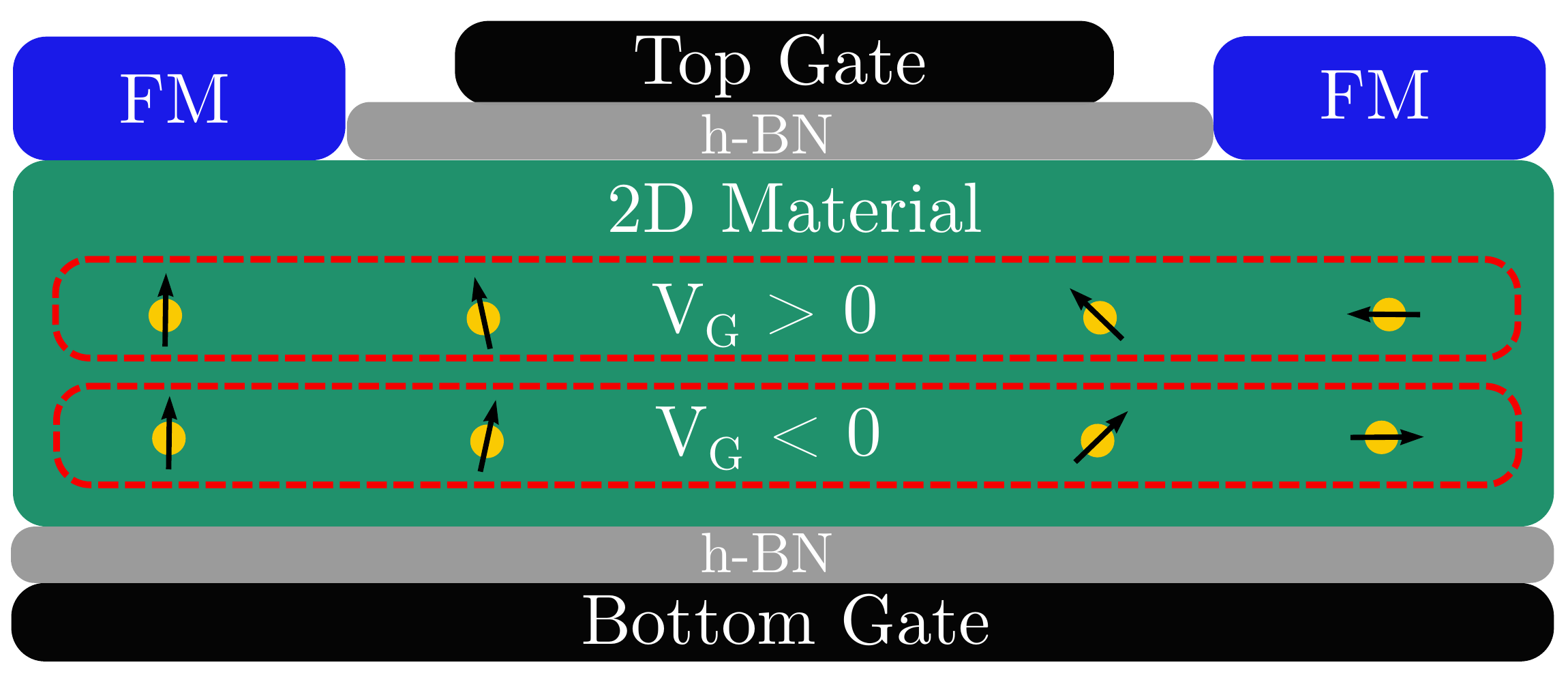}
  \includegraphics[align=c,width=0.49\textwidth]{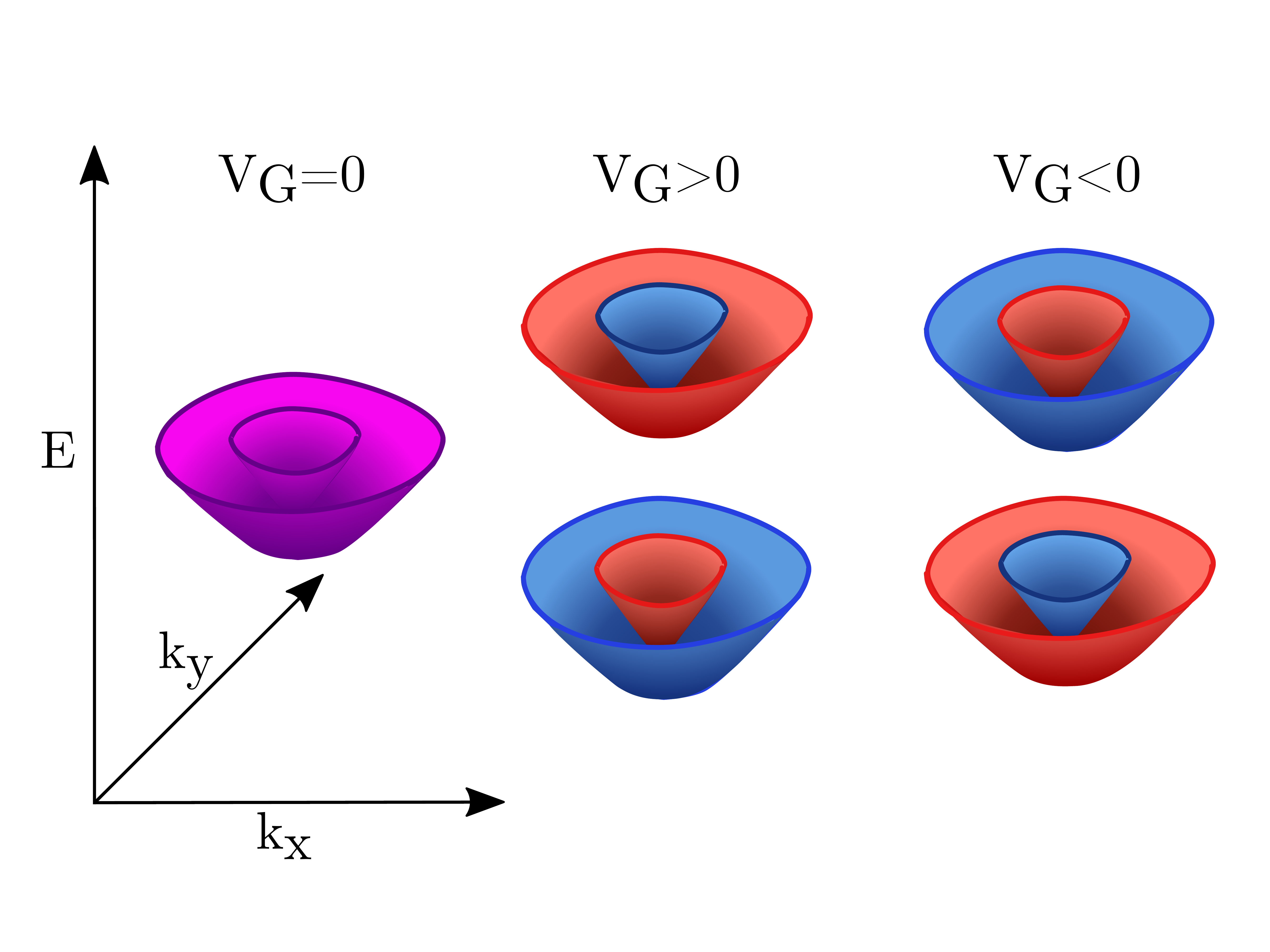}
  \caption{Left panel: schematics of an advanced R-2 Datta-Das spin field-effect transistor (spinFET) with 2D materials, where ferromagnetic (FM) contacts are marked in blue, electrical gates in black and the separating dielectric, made of hexagonal boron nitride (h-BN), in grey. Spin currents are injected through a ferromagnetic source and undergo spin precession in the 2D material before being collected by the ferromagnetic sink. In a standard Datta-Das spinFET a gate voltage would manipulate the spin precession by controling the strength of the Rashba spin-orbit coupling (SOC), while in this setup the gate voltage controls the direction of spin precession. Right panel: the advanced R-2 Datta-Das spinFET can be realized through centrosymmetric 2D materials with local dipoles (R-2 materials), where two-fold degenerate Rashba bands (magenta) can be split in energy by a gate potential. The two sets of bands have opposite spin textures (in-plane spin expectation values, with blue and red indicating clockwise and anticlockwise rotation respectively, see supporting information) and the voltage sign determines which one is populated by doping.}
  \label{Fig1}
\end{figure*}
In these R-2 materials, such as \ce{LaOBiS2} \cite{LaOBiS2NanoLetter,zhang_hidden_2014}, bands are still doubly degenerate owing to the inversion-symmetric space group, but they are composed by two branches with opposite polarization summing up to zero net polarization\cite{zhang_hidden_2014}.
The presence of site dipole fields is responsible for creating the typical Rashba-split band structure (see Fig.\ref{Fig1}b), with two crossing parabolas and the helical spin texture, but in the R-2 materials the electronic states corresponding to each degenerate branch are localized on different regions of the material in real space (as shown in Fig. \ref{Fig2}).
This spin separation in van-der-Waals (vdW) materials \cite{yao_direct_2017,lee_unveiling_2020} has also been named spin-layer locking (SLL).
As discussed in Ref. \citenum{LaOBiS2NanoLetter}, the degeneracy can be lifted by applying a relatively weak external field (on the order of $1$ V/nm, commonly achieved in field-effect devices), hence giving the possibility to select just one of the two \emph{channels}. Those channels are localized in different regions of the material and exhibit \emph{opposite spin precessions}.
This is one of the gate-controlled aspects of the device operation. A second, which has not been considered in previous theoretical work, is the electrostatic doping of the system, needed to add free carriers in the semiconductor and to bring the Fermi level into the bands with the appropriate spin-texture.

In this work, we first introduce lutetium oxide iodide, LuIO, a novel exfoliable monolayer that is a promising R-2 material to realize the advanced spinFET introduced in Ref. \citenum{LaOBiS2NanoLetter}.
Then, we discuss how to operate the spinFET through electrical gates by explicitly simulating two key mechanisms: electrostatic doping and the splitting of the spin channels' bands.
A finite electrostatic doping is necessary to place the Fermi level within a region of the bands with an appropriate spin texture.
We show that this doping interferes non-trivially with the ability to split the channels' bands and thus switch the device.
According to our knowledge, this is the first time that field-effects are fully included in 2D R-2 materials.

%\section{LuIO: an easily-exfoliable R-2 material with large Rashba effect}
\textbf{LuIO: an easily-exfoliable R-2 material with large Rashba effect.}
Lutetium oxide iodide, LuIO, can be obtained as a by-product of the reaction of lutetium metal, rhenium powder and lutetium triiodide, LuI$_3$, in a sealed tantalum container \cite{exp_LuIO_2007}.
LuIO crystallizes in the tetragonal PbFCl structure-type (matlockite), where Lu, O and I are located on sites with 4mm, 4m2 and 4mm symmetry, respectively \cite{exp_LuIO_2007}.
LuIO is a layered crystal with AA stacking \cite{exp_LuIO_2007} and very low binding energy, as calculated using non-local vdW functionals in Ref. \citenum{MaterialsCloud}.
The vdW-DF2 functional\cite{lee_df2_09}  with  C09  exchange  (DF2-C09)\cite{cooper_c09_10} yields a binding energy of $E_b^{DF2-C09}=15.5 \hspace{1pt}\text{ meV}/\si{\angstrom}^{-2}$, that becomes $E_b^{rVV10}=22 \hspace{1pt}\text{ meV}/\si{\angstrom}^{-2}$ for the rVV10 functional\cite{vydrov_vv10_09,sabatini_rvv10_13}.
For reference, the binding energy of graphene is $E_b^{DF2-C09}=20 \hspace{1pt}\text{ meV}/\si{\angstrom}^{-2}$ and $E_b^{rVV10}=26 \hspace{1pt}\text{ meV}/\si{\angstrom}^{-2}$, making LuIO as an easily exfoliable material\cite{MaterialsCloud}.

The crystal structure of monolayer LuIO is shown in Fig.\ref{Fig2} and can be discussed in terms of three planes.
The inner flat plane contains a square lattice of oxygen atoms, rotated by $\pi/4$ with respect to the unit cell (the O-O distance is smaller than the the lattice constant by a factor $1/\sqrt{2}$).
The oxygen layer is sandwiched by two planar square lattices made of lutetium atoms that are chemically bonded with the oxygen atoms.
The two lutetium layers are displaced one from the other by $(1/2,1/2)$ in reduced coordinates.
The two outer layers of iodine atoms follow the pattern of the closest lutetium atoms, the two being chemically bonded, but again with a relative shift of $(1/2,1/2)$.
The strong dipole field between the outer iodine and the inner lutetium atoms is responsible for the strong Rashba effect, as it can be observed in the potential drop reported in Fig.\ref{Fig2}.

\begin{figure*}
  \includegraphics[width=0.47\textwidth]{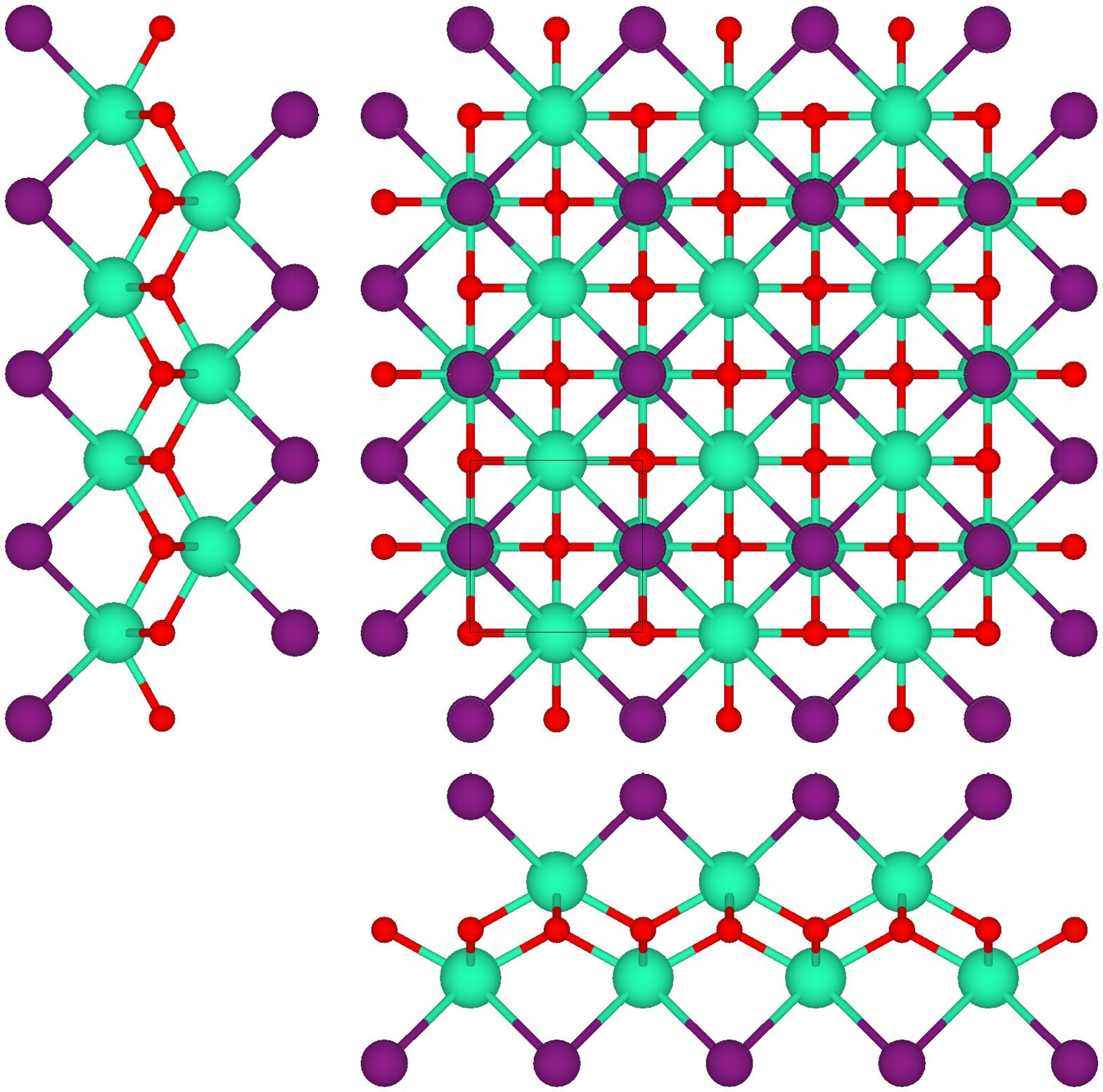}
  \includegraphics[width=0.49\textwidth]{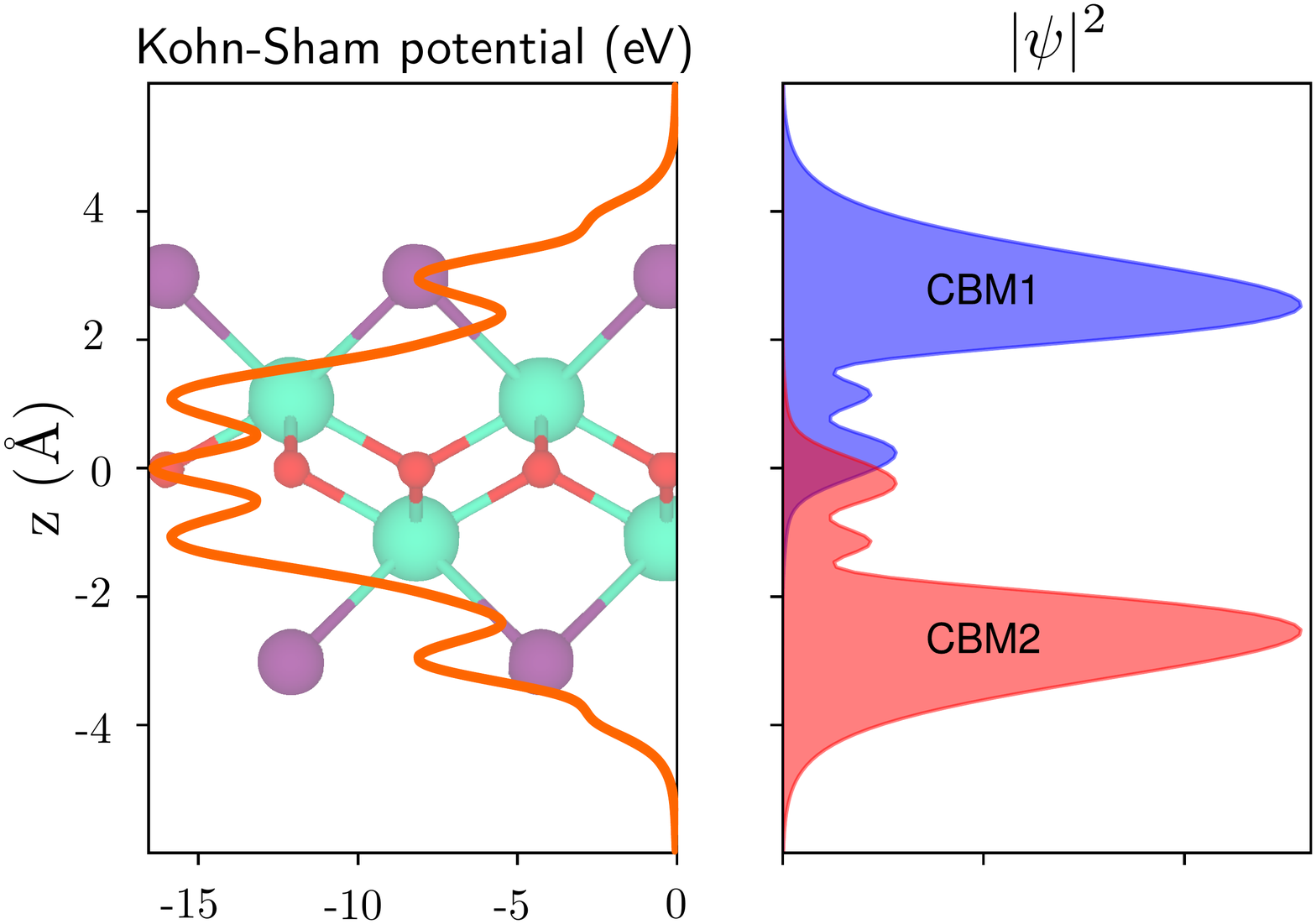}
  \caption{Left: crystal structure of LuIO (top and lateral views), where Lu atoms are depicted in acquamarine, I atoms in purple and O in red. Center: planar average of the electrostatic Kohn-Sham potential felt by a test charge, plotted along the out-of-plane direction. Right: planar average of the charge density calculated for the conduction band minima. The two densities, depicted in red and blue respectively, are localized in the two different halves of the material that host I atoms.}
  \label{Fig2}
\end{figure*}
\begin{figure}
  \includegraphics[width=0.52\textwidth]{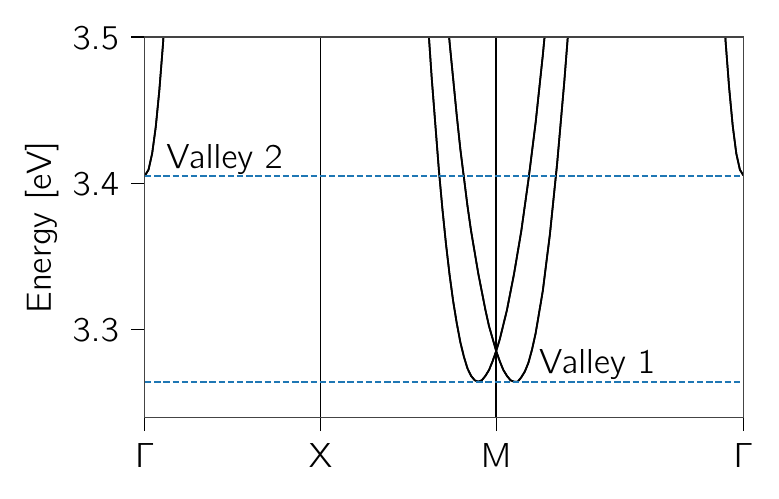}
  \caption{Band structure of the first conduction bands of monolayer LuIO, obtained using density-functional theory with the PBE functional and including spin-orbit coupling. Two valleys very close in energy are present, where the lowest valley has the characteristic conical shape due to the Rashba effect.}
  \label{Fig3}
\end{figure}

We compute the band structure of monolayer LuIO using density-functional theory (DFT, details in Methods); the first conduction bands are shown in Fig.\ref{Fig3}.
The full band structure including also the valence band is reported in Fig. S1 of the supporting information.
Within our computational framework, monolayer LuIO is an insulator with a band gap of $3.18$ eV and two valleys in the conduction band.
The lowest valley is centered around M and it is affected by Rashba SOC, visible in the typical conical shape.
The second valley is parabolic and located at the $\Gamma$ point, with the bottom being only $0.14$ eV higher in energy than the bottom of the lowest valley.
The effect of Rashba SOC is very strong on the first valley and it is quantified by the Rashba parameters $k_R = 0.08$ $\si{\angstrom}^{-1}$, $E_R = 0.07$ eV and $\alpha_R = \frac{E_R}{k_R} = 0.9 \text{ } \si{eV}\si{\angstrom} $ (we define $\alpha_R$ as in Ref. \citenum{LaOBiS2NanoLetter}).
Hence, monolayer LuIO hosts one of the strongest Rashba effects among all known 2D materials \cite{yao_direct_2017} and a $k_R$ much larger than LaOBiS$_2$ \cite{LaOBiS2NanoLetter}.
While LaOBiS$_2$ is actually made of three ionically bonded sublayers  \cite{LaOBiS2NanoLetter}, namely BiS$_2^-$/(LaO)$_2^{2+}$/BiS$_2^-$, LuIO is composed of a single covalently-bonded monolayer (see Fig.\ref{Fig2}).

\textbf{Electrostatics of the ``spin-channel switch''.}
\begin{figure*}
  \includegraphics[width=0.55\textwidth]{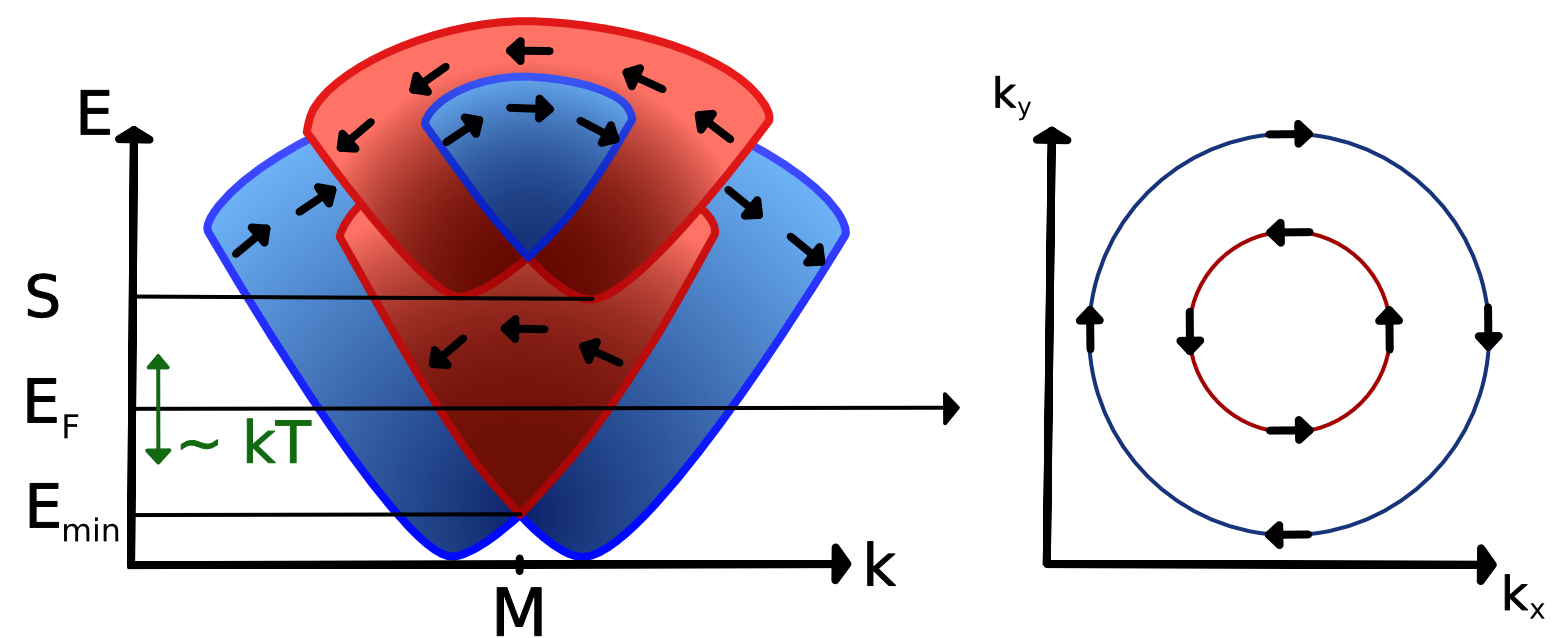}
  \includegraphics[width=0.4\textwidth]{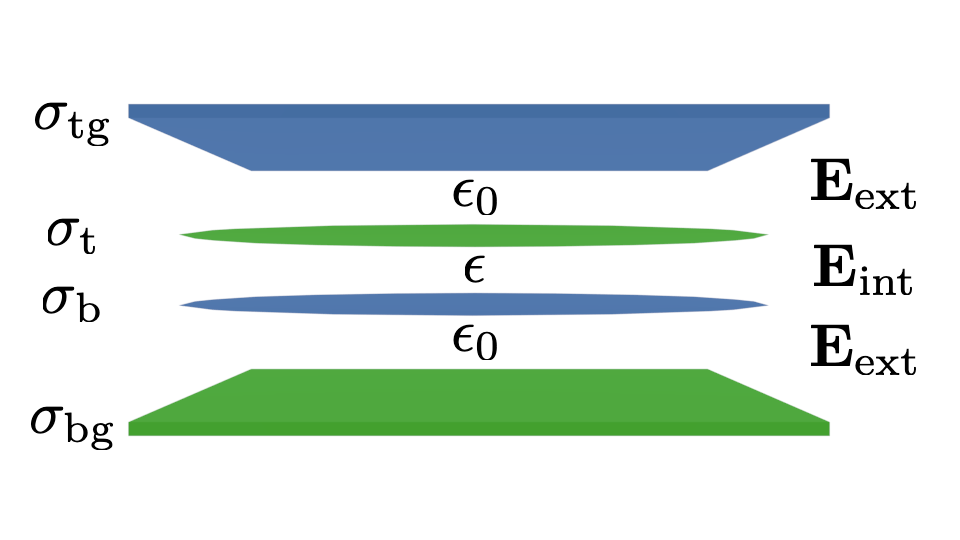}
  \caption{The left panel describes the operation of the device from the electronic structure point of view. The Fermi level is inside the band with Rashba spin-texture. The spin current flows in an energy window of $kT$ around the Fermi level. The energy splitting of the spin channel $S$ should be such that the top channel is not occupied.
  The right panel represents the electrostatic model of the device, made of four charged planes: two gates and two spin-channels. The gates allow one to control which channel the charges go through. In this picture, blue and green represent negative and positive charges, respectively, such that the external electric field generated by the gates points in the upward direction. This is the situation assumed in the text, although the opposite polarity follows the same equations with the appropriate sign changes.}
  \label{fig:channel-switch}
\end{figure*}
In the following, we develop a simple electrostatic model for this material in a double-gate FET setup.
The main quantities introduced here are described in Fig. \ref{fig:channel-switch}.
We account for two basic aspects of the operation: i) the out-of-plane electric field to break the degeneracy of the channels and ii) an imbalance of absolute charge between the gates to dope the material.
This second aspect was not treated in Ref. \citenum{LaOBiS2NanoLetter}, although it is essential to the device operation, since the additional doping electrons carry the spin current.

As shown in Fig. \ref{fig:channel-switch}, two charged planes play the role of the top and bottom gates. The sum of their planar charge density $\sigma_{tg}+\sigma_{bg}$ neutralizes the planar charge density of the material $\sigma = \sigma_{t}+\sigma_{b}$. The material is modeled as two separate conductive channels represented by two charged planes. The screening effect in the material is accounted for by a dielectric medium of dielectric constant $\epsilon$.\\

For spin-FET applications, we have to fix the doping density, hence the sum of the gate charges. The difference $\Delta \sigma = \sigma_{bg}-\sigma_{tg}$ is the degree of freedom that can be tuned to generate an out-of-plane external field.
The external electric field generated by the two gates (in vacuum) is given by electrostatics:
\begin{align}
\Ex  = \frac{\sigma_{\rm{bg}} - \sigma_{tg}}{2\epsilon_0} = \frac{\Delta \sigma}{2\epsilon_0}.
\end{align}
It controls in which conducting channel the doping charges will accumulate.
Added charges will tend to go towards the gate with opposite polarity. In the band structure, the otherwise degenerate conduction bands split and the electrons accumulate in the lowest one. The energy of the splitting $S$ corresponds to the potential drop between the two charged planes representing the conductive channels, and is proportional to the electric field $\Ei$ inside the material.
The proportionality factor is related to the distance between the conductive channels. This quantity depends slightly on the electrostatic setup as the structures are relaxed. However, we assume this dependency to be relatively mild with respect to the other effects discussed in the following and note the proportionality constant $\beta$.

The internal electric field $\Ei$ is found by summing the external electric field screened by the (neutral) material plus a contribution from the material's charged planes:
\begin{align}\label{eq:S}
S = \beta \Ei = \beta \left( \frac{\Delta \sigma}{2\epsilon} + \frac{\sigma_{b} - \sigma_{t}}{2\epsilon} \right).
\end{align}

To maximize the spin-polarity of the carrier distribution, the Fermi level $E_f$ should be in only one of the bands, with some margin corresponding to the smearing of the Fermi-Dirac distribution at room temperature. Thus, we need a large enough energy separation $S$ between the bands.
The two parameters ($E_f$ and $S$) are essential for the operation of this device and they can be controlled with the gate parameters $\sigma$ and $\Delta\sigma$, respectively.
In the following we perform an extensive ab initio study and propose a method to choose the controllable external parameters $\sigma$ and  $\Delta \sigma$ in order to achieve ideal operation conditions of such spintronic device. We perform calculations with the material placed between the gates in vacuum.
A more realistic model would include gate dielectrics between the gates and the material. The corresponding dielectric constant would rescale the parameter $\beta$, as could, potentially, other experimental parameters. Here we establish a basic formalism that can be easily adapted to specific experimental configurations.

\begin{figure*}
  \includegraphics[width=0.49\textwidth]{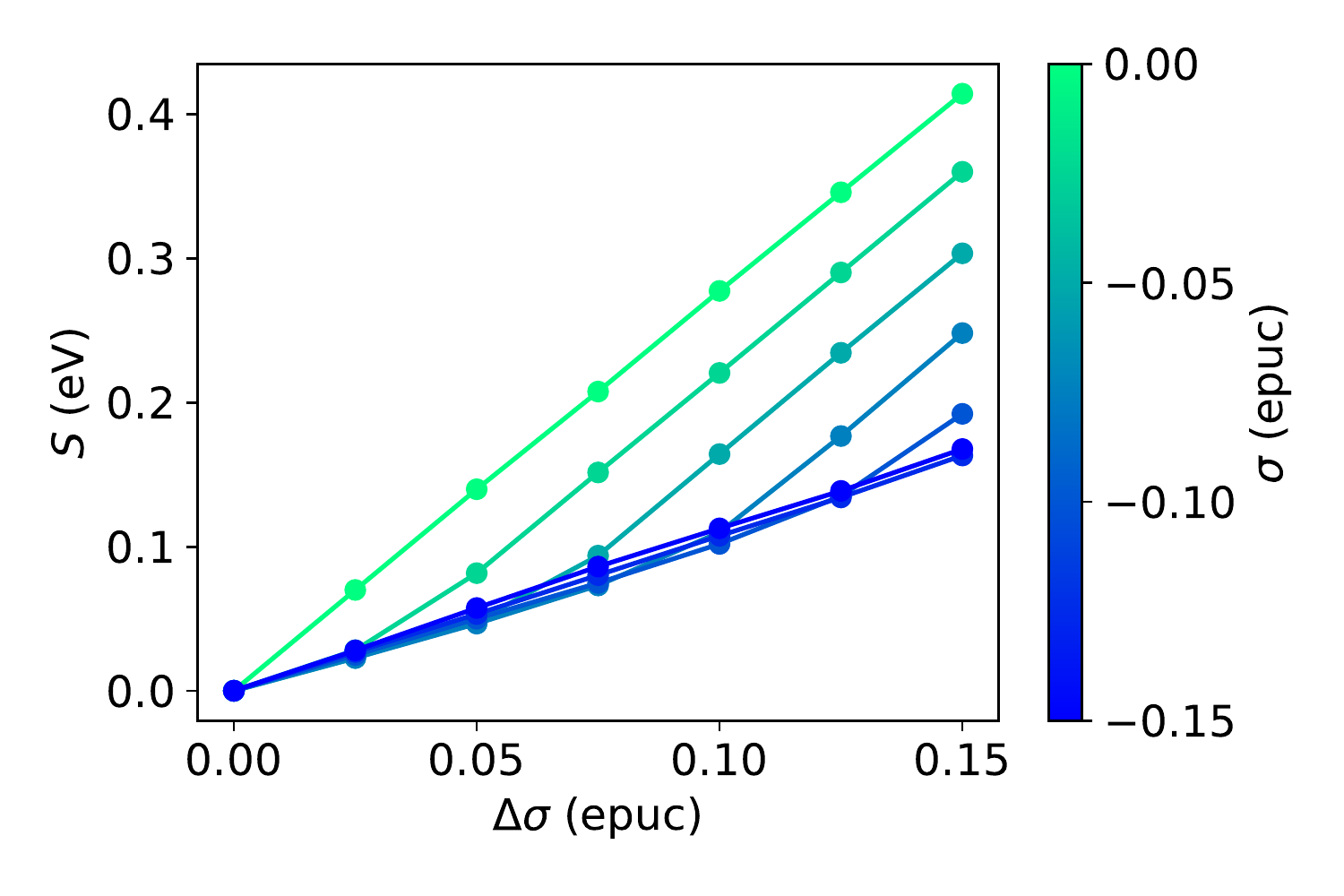}
  \includegraphics[width=0.49\textwidth]{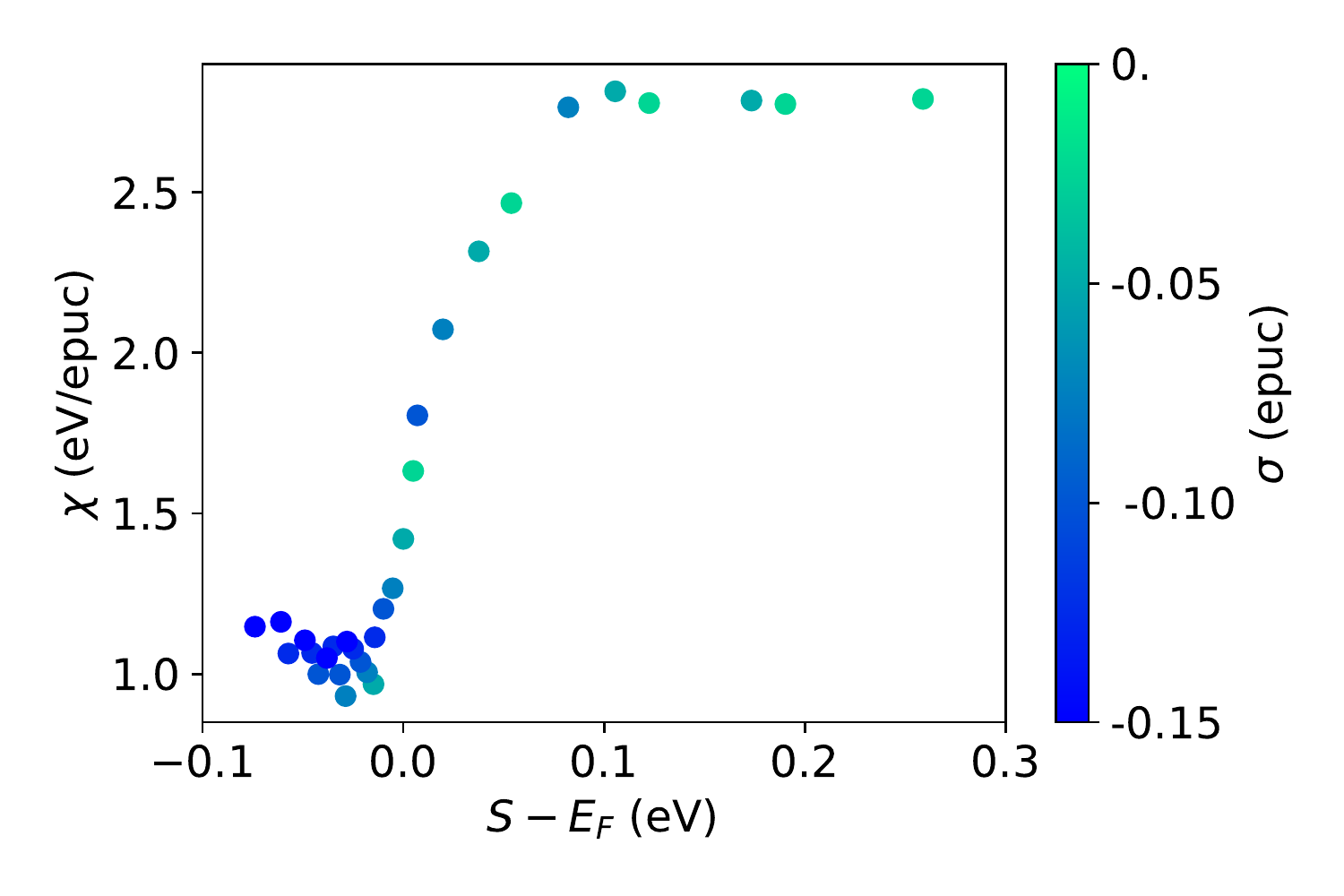}
  \caption{
  Left pannel: conduction band energy splitting $S$ of \ce{LuIO} as a function of the gate charge difference $\Delta \sigma$, at different doping charge $\sigma$.
  The ``epuc'' units stands for "elementary charge per unit cell". For n-type doping, the charge of the material $\sigma$ is negative.
  $S$ increases with $\Delta \sigma$ at a different rate depending on the regime, as described in the text.
  Right panel: Suceptibility as a function of the relative position of the second channel's band ($S$) and the Fermi level ($E_F$), see Fig. \ref{fig:channel-switch}, for different doping conditions.
  The transition between the high and low susceptibility regimes is clearly driven by the $S-E_f$ parameters, that is the occupation of the second channel's band.}
  \label{fig:DFTresults}
\end{figure*}

\textbf{DFT results.}
The DFT setup used to obtain the following results is described in the Methods section. It includes top and bottom gates with arbitrary charge. As discussed in the supporting information, the material is dynamically stable in this setup.
In the left panel of Fig. \ref{fig:DFTresults}, we plot the splitting of the bands $S$ as a function of $\Delta \sigma$ (itself proportional to the external electric field $\Ex$).
The slope of those curves represents how easy it is to split the bands with an external electric field. We will refer to it as the susceptibility $\chi$.
We observe that $S$ increases linearly with $\Delta \sigma$, with two different susceptibilities corresponding to two different regimes.

The high susceptibility (large slope) regime is accessed when the doping is low enough and the external electric field is high enough that only one band  is occupied, corresponding for example to the bottom channel. Indeed, modifying Eq \ref{eq:S} according to this situation we have :
\begin{align} \label{eq:reg1}
  S = \beta\left(\frac{\Delta \sigma}{2\epsilon} +\frac{\sigma_{\rm{b}}}{2\epsilon} \right).
\end{align}
Note that for n-type doping of the bottom channel, $\sigma_b$ is negative.
The susceptibility is $\chi =  \frac{\beta}{2\epsilon}$ and the curve of the left pannel of Fig. \ref{fig:DFTresults} would intercept the $y$-axis
at a finite value $S(\Delta \sigma = 0) = \beta \frac{\sigma_{\rm{b}}}{2\epsilon}$, although one cannot reach this point in practice since the external electric field would not be large enough anymore.

In the low susceptibility regime, the doping is high enough for the Fermi level to be above the bottom of the second band. We introduce the threshold charge $\sigma_{\rm{thr}}$ at which the second band starts to be filled, i.e. when the Fermi level with respect to the bottom of the first band is equal to $S$. Assuming a constant density of states $D$, which is reasonable in 2D materials and exact in the limit of a non-interacting 2D electron gas, we can relate the threshold charge to the splitting as such:
\begin{align} \label{eq:sthr}
    \sigma_{\rm{thr}} = - e D S,
\end{align}
where $e$ is the elementary charge. Assuming the same constant density of states $D$ for both bands, any charge in addition to $\sigma_{\rm{thr}}$ will distribute equally over both conductive channels.
The charge difference between the channels thus saturates at $\sigma_{\rm{thr}}$, and so does the materials' contribution to the internal electric field. We can write Eq. \ref{eq:S} as:
\begin{align}
S &= \beta \left(  \frac{\Delta \sigma}{2\epsilon} + \frac{\sigma_{\rm{thr}}}{2\epsilon} \right).
\end{align}
Considering that $\sigma_{\rm{thr}} = - e D S$ depends on the splitting, we finally have:
\begin{align} \label{eq:reg2}
S &=  \frac{\beta \Delta \sigma}{2\epsilon+ eD \beta} .
\end{align}
The susceptibility is now $\chi = \frac{\beta}{2\epsilon+ eD \beta}$.
Compared with the first regime, it is smaller by a factor $1+\frac{eD \beta}{2\epsilon}$, and the curves now go through the origin.

We plot the susceptibility $\chi$ in the right panel of Fig. \ref{fig:DFTresults}.  The two regimes correspond to the regions where the susceptibility is constant, with a transition in between.

Using the high susceptibility regime value of
$\chi =  \frac{|\beta|}{2\epsilon}$ and $eD = (\frac{1}{\chi_{L}} - \frac{1}{\chi_{H}}) = 0.55$ epuc/eV (where epuc stands for electrons per unit cell), the parameters of the model are known and we can estimate values of the doping and the charge difference between the gate corresponding to certain operating conditions.

Setting the ideal conditions to be $S>kT$ and $|\sigma|=\sigma_{\rm{thr}}/2$, we obtain:

\begin{align}
\Delta \sigma  &> \left(\frac{eD}{2}+\frac{2\epsilon}{\beta} \right) kT , \\
    |\sigma| & = \frac{eD\beta}{4\epsilon+eD\beta} \Delta \sigma = \frac{eD/2}{2\epsilon/\beta + eD/2}\Delta \sigma .
\end{align}

At room temperature, and for the current setup with vacuum, the above condition translates into $\Delta \sigma > 0.016$ electrons per unit cell and $|\sigma| = 0.43 \Delta\sigma$.
Another condition for the spin-FET operation, not related to the electrostatics but to the spin texture, is that the Fermi level is higher than the energy $E_{min} \approx 17.5$ meV defined on Fig. \ref{fig:channel-switch}. This leads to $|\sigma|> eD E_{min} = 0.01$ electrons per unit cell, or an electron density around $n = 7 \ 10^{12}$ cm$^{-2}$. In LuIO, this happens to be more constraining than the purely electrostatic considerations above. The minimum gate charge difference is affected accordingly: $\Delta \sigma > 0.023$ electrons per unit cell. In any case, and for future application of this model to other materials, one must keep in mind that both conditions must be satisfied.

%\bibliographystyle{plain}

%\section{Conclusions}
In conclusion, we propose monolayer LuIO as a novel easily-exfoliable centrosymmetric material with local dipoles, displaying one of the strongest Rashba spin-orbit couplings among 2D materials, accompanied  by the spin-layer locking effect, where two separate sublayers exhibit degenerate and opposite spin textures. First, we show that monolayer LuIO is an ideal candidate to implement an advanced Datta-Das spinFET based on the spin-layer locking mechanism, in which the two spin textures can be selectively populated by controlling the gate voltage in a double-gate field-effect setup. The strong Rashba effect allows to construct extremely small devices with a very short channel length of 1 nm.  Second, we perform in-depth analysis of the device electrostatics through DFT simulations by explicitly including, for the first time, the effect of doping and external electric fields. We also develop a tailored analytical model to understand the interplay between the two key device parameters: the energy splitting of the two spin channels and the amount of electron doping. Those two aspects are controlled via the difference and the sum of the gate charges in a double gate setup, respectively. We show that the amount of splitting between the two spin channels due to the gate strongly depends on doping and decreases when both spin channels start to be occupied. This phenomenon is fully captured by our analytical model that allows to naturally determine the optimal gate-control parameters for device operation. We emphasize that our findings are not limited to LuIO but they apply to the entire class of 2D centrosymmetric materials with local dipoles, providing an insightful method to engineer spin-layer-locking spinFETs.

\section{Methods}
\textbf{Ab-initio simulations.}
DFT simulations are performed using the Quantum ESPRESSO distribution \cite{Giannozzi2009,Giannozzi2017}, adopting the SSSP efficiency library and cutoffs v1.1  \cite{Prandini2018,emine_pslib_14} for structural relaxation, phonon calculations and field-effect simulations. The PBE functional \cite{perdew_pbe_96} is used despite the presence of f-electrons (see supporting information).
Band structure calculations include the effect of SOC and are computed using the fully-relativistic PseudoDojo library \cite{dojo_paper_18,ONCVPSP}, with 100 Ry and 400 Ry of cutoff on the wavefunction and charge density respectively.

\textbf{Field-effect calculations.}
Field-effect calculations are carried out by applying 2D periodic boundary conditions and simulating electrical gates as described in \cite{FETcode},  using a modified version of Quantum ESPRESSO that is available at https://gitlab.com/tsohier/qe-2D-FET. The gates and the material are separated by potential barriers to hold the material in place, thus emulating the purely mechanical role played by the dielectric or encapsulator in real devices.
For every electrostatic configuration, determined by the charge of the material and the gate-charge difference, we optimize the atomic positions and compute the band structure. It is important to perform those calculations with a rather fine momentum grid and small smearing to emulate the Fermi-Dirac occupation at room temperature, hence we use $40 \times 40 \times 1$ k-point grids and a Marzari-Vanderbilt \cite{mv_smearing_99} electronic smearing of $0.002$ Ry.

\section{Acknowledgements}
The results of this research have been partially achieved using the DECI resource ARCHER UK National Supercomputing Service with support from the PRACE aisbl. Simulation time was also awarded by PRACE (project id. 2020225411) on MareNostrum at Barcelona Supercomputing Center - Centro Nacional de Supercomputación (The Spanish National Supercomputing Center) and on MARCONI at CINECA Italy (project id. 2016163963).
R.Z., A.M., N.M. and T.S. acknowledges support from NCCR MARVEL funded by the Swiss National Science Foundation, R.Z. was supported by the NCCR MARVEL INSPIRE Potentials fellowship. T.S. acknowledges support from the University of Liege under the Special Funds for Research, IPD-STEMA Programme.

\section{Supporting Information}

\subsection{Band structure with spin-orbit coupling}
\begin{figure}
  \includegraphics[width=0.49\textwidth]{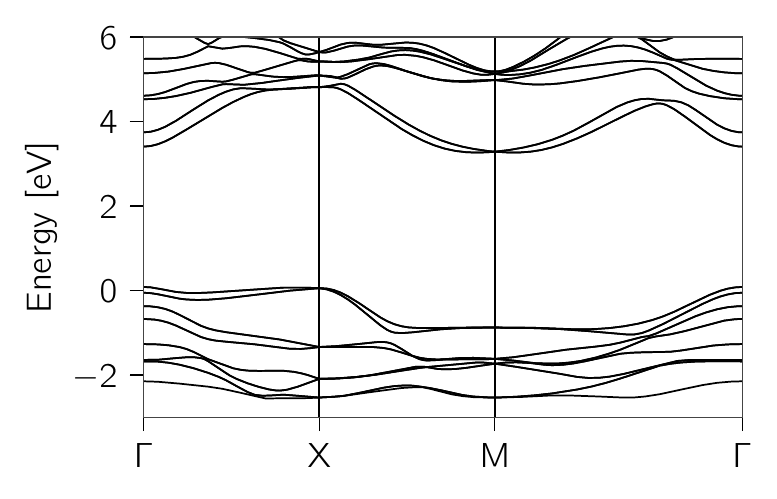}
  \caption{Band structure of monolayer LuIO at the DFT-PBE level with SOC.}
  \label{SupplFig1}
\end{figure}

Fig. \ref{SupplFig1} shows the band structure of LuIO with spin-orbit interactions.

\subsection{Phonons and stability}
\begin{figure}
    \centering
    \includegraphics[width=0.25\textwidth]{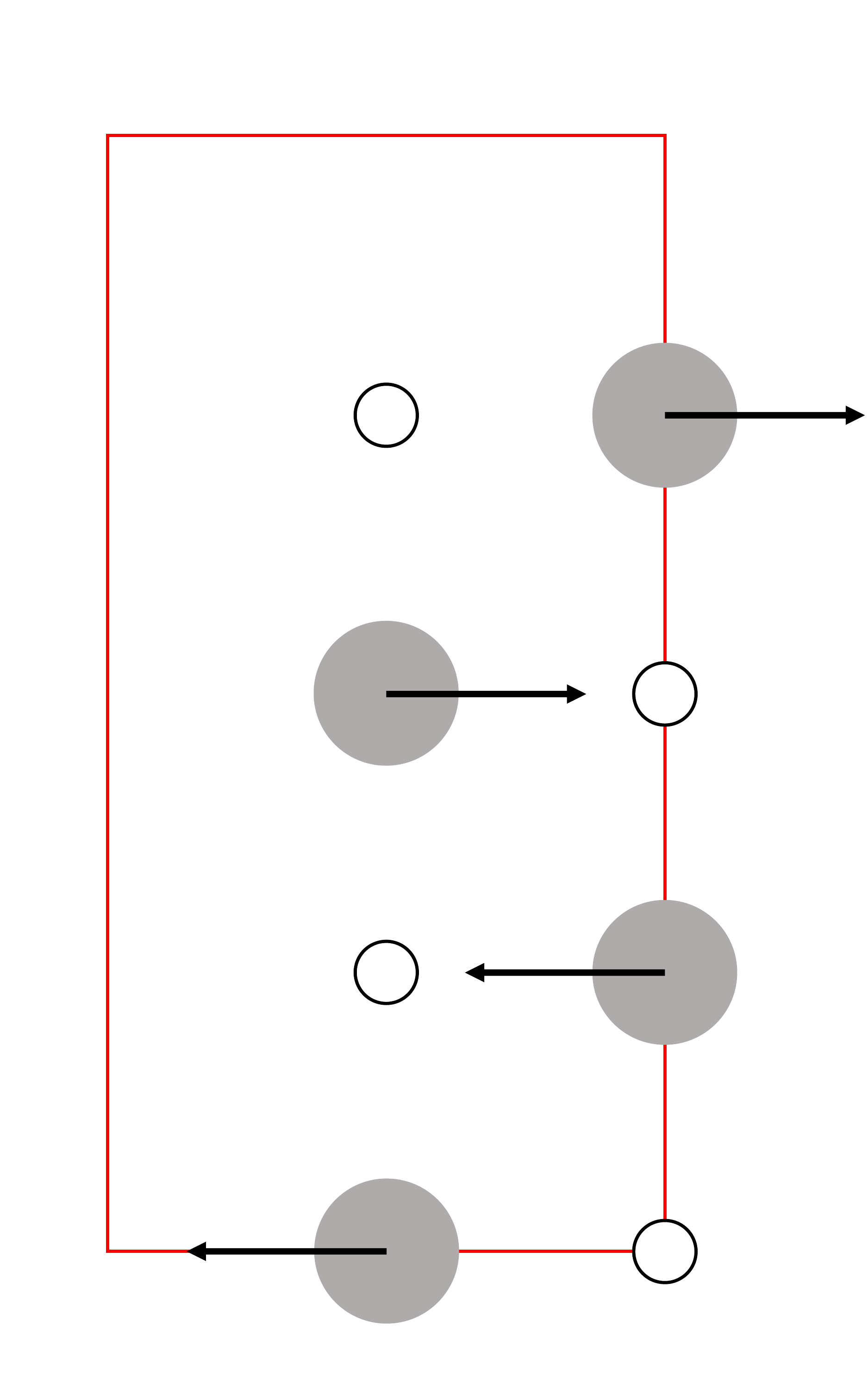}
    \caption{Schematic diagram of phonon instability at X-point. Red: $1\times2$ supercell. Grey circle: iodine atoms at high-symmetry positions. Solid lines: oxygen atoms. Lutetium atoms are neglected, for they overlap with the iodine atoms on x-y plane. Arrows: direction of displacement of iodine atoms.}
    \label{fig:phonon}
\end{figure}

The phonon spectrum of LuIO, as shown in the 2D structures and layered materials database \cite{MaterialsCloud} available on the Materials Cloud \cite{MC2D}, indicates the instability of two phonon modes at X-point. These phonon modes corresponds to transverse displacements of the iodine atoms with respect to the wave vector, Fig. \ref{fig:phonon}.
However, those phonon modes are positive for the monolayer ILuO sandwiched between potential barriers that are reasonably close to the material, as used in our calculations to emulate the electronic and ionic charge density of the encapsulator or dielectric material separating the layer from the gates in experiments. More specifically, barriers $2.66$ \AA\ away from the iodine atoms are sufficient to stabilize the high-symmetry structure. Given that the distance in the z-direction between two iodine atoms from two successive layers is $3.92$ \AA, we argue that FET setup in the experiment would be able to stabilize the negative phonon modes found in the suspended material.

\subsection{The role of $f$-electrons}
\begin{figure}
  \includegraphics[width=0.49\textwidth]{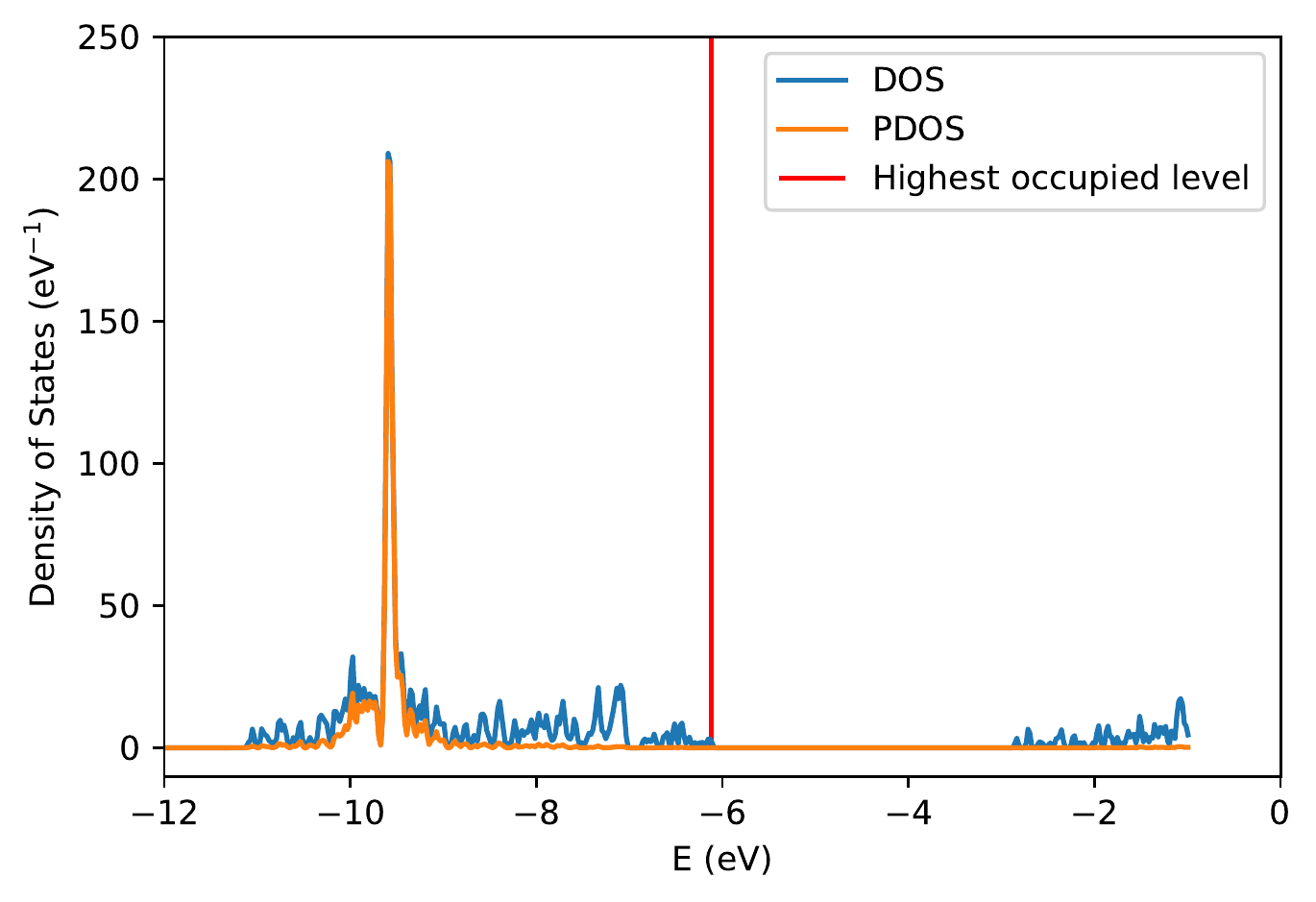}
  \caption{Total density of states (blue line) and projected density of states on $f$-orbitals (orange line) of monolayer LuIO at the DFT-PBE level without SOC. The red line displays the Fermi level.}
  \label{SupplFig2}
\end{figure}

The electronic configuration of Lutetium includes a fully-occupied shell of $f$-orbitals, treated as valence electrons in the pseudopotential \cite{Prandini2018}. It is well known that $f$-electrons might not be correctly described by semi-local DFT (such as DFT-PBE) and they may require beyond-DFT methods to obtain a qualitatively-correct electronic structure. However, Lu contains a \emph{full} $f$-shell and so the $f$-orbitals are all quite low in energy, on the order of 4 eV below the top of the valence band already at the PBE level, as shown in Fig. \ref{SupplFig2}. Hence, in LuIO $f$-electrons do not affect the transport properties and they are not involved in the discussion of the electronic structure above the Fermi level that is present in the main text.

\subsection{Spin texture}

\begin{figure*}
  \includegraphics[width = \textwidth]{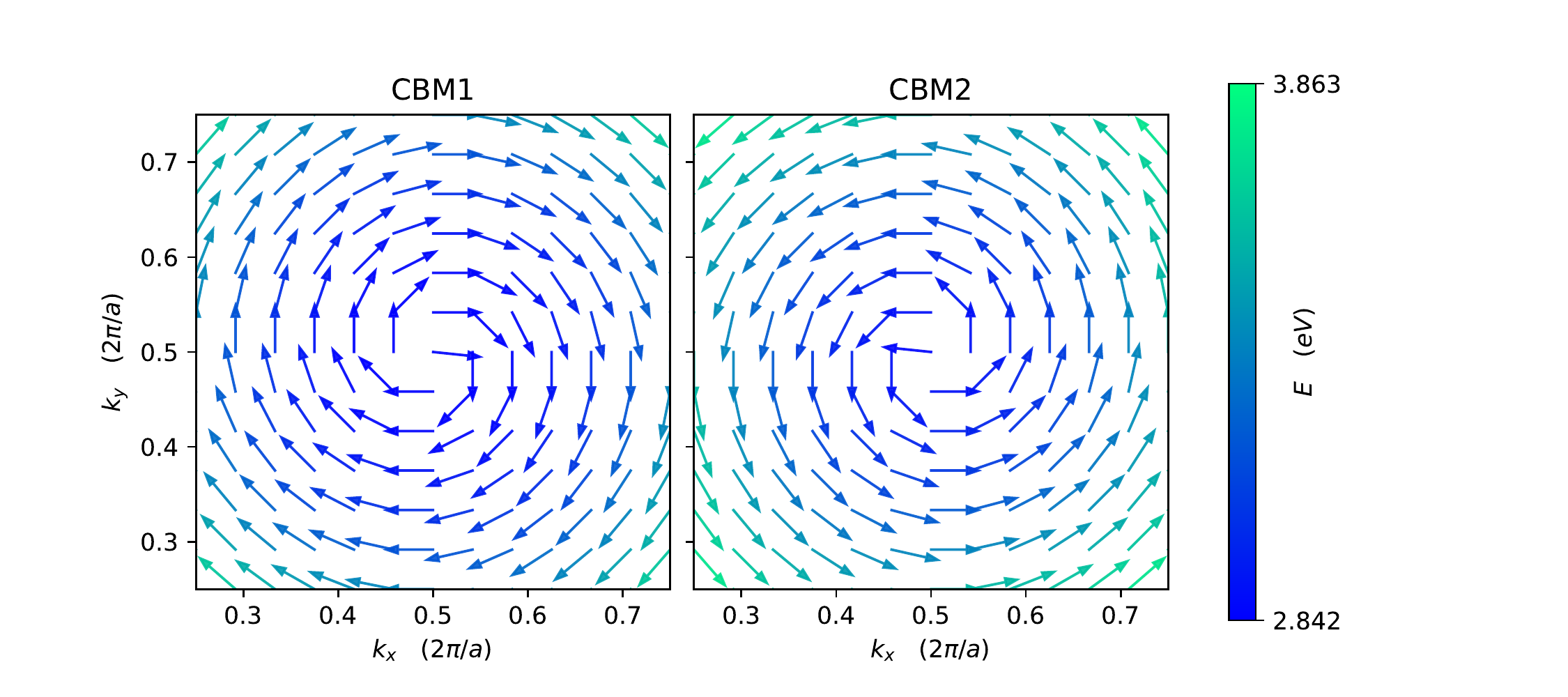}
  \caption{Spin texture of conduction band minima around $M$ in the vertical electric field. The energy in the colorbar takes the valence band maximum as reference. The x and y spin projections of each state can be read from the length of x and y components of the arrows with a scaling factor of 10. }
  \label{SupplFig3}
\end{figure*}
For the spin texture of the conduction band minima in Fig. \ref{SupplFig3}, the electric field, which breaks the inversion symmetry, is implemented by imposing a bottom gate charge of $0.03 \enspace\textrm{epuc}$. The spin projection components are oriented within the x-y plane, displaying a typical Rashba type spin texture. Along with the band structure in Fig. \ref{SupplFig1}, the spin texture justifies our schematic diagrams in the main text.

\bibliography{biblio}
\end{document}